\documentclass[twocolumn]{svjour3}

\smartqed  

\usepackage{verbatim}

\usepackage{underscore}
\usepackage{cite}

\usepackage{amsmath}
\usepackage{amssymb}
\usepackage{stmaryrd} 
\usepackage{wasysym}
\usepackage{mathtools} 
\usepackage{bm}
\usepackage{marvosym}

\usepackage{pifont}
\usepackage{multicol}
\usepackage{hyperref}

\usepackage{multicol}

\usepackage[linesnumbered]{algorithm2e}

\usepackage{listings}
\usepackage{xcolor}
\lstdefinestyle{verilog-style}
{
	language=Verilog,
	basicstyle=\small\ttfamily,
	keywordstyle=\color{vblue},
	identifierstyle=\color{black},
	commentstyle=\color{vgreen},
	numbers=left,
	numberstyle=\tiny\color{black},
	numbersep=10pt,
	tabsize=8,
	moredelim=*[s][\colorIndex]{[}{]},
	literate=*{:}{:}1
}
\makeatletter
\newcommand*\@lbracket{[}
\newcommand*\@rbracket{]}
\newcommand*\@colon{:}
\newcommand*\colorIndex{%
	\edef\@temp{\the\lst@token}%
	\ifx\@temp\@lbracket \color{black}%
	\else\ifx\@temp\@rbracket \color{black}%
	\else\ifx\@temp\@colon \color{black}%
	\else \color{vorange}%
	\fi\fi\fi
}

\usepackage{tikz}
\usetikzlibrary{arrows,automata,positioning,shapes}

\usepackage{pgfplots}
\pgfplotsset{compat=1.12}
\definecolor{vgreen}{RGB}{104,180,104}
\definecolor{vblue}{RGB}{49,49,255}
\definecolor{vorange}{RGB}{255,143,102}

\usepackage{ltltikz}
\usepackage{ltl}

\usepackage[utf8]{inputenc}
\usepgfplotslibrary{groupplots}
\pgfplotsset{compat=newest}

\newcommand{\set}[1]{\{#1\}}

\newcommand{\card}[1]{{|#1|}}
\newcommand{\ldot}{\mathpunct{.}}

\newcommand{\Tau}{\mathcal{T}}
\newcommand{\dom}{\mathrm{dom}}

\newcommand{\hlang}{\mathcal{H}}

\newcommand{\ltl}{\text{LTL}}
\newcommand{\hyperltl}{\text{HyperLTL}}
\newcommand{\secltl}{\text{SecLTL}}

\newcommand{\ap}{\text{AP}}
\renewcommand{\models}{\vDash}
\newcommand{\nmodels}{\nvDash}
\newcommand{\traces}{\mathit{TR}}

\newcommand{\G}{\Globally}

\newcommand{\W}{\WUntil}

\newcommand{\pathassign}{\Pi}
\newcommand{\pathvars}{\mathcal{V}}
\newcommand{\pathassignfin}{\Pi_\mathit{fin}}

\newcommand{\monitor}{\mathcal{M}}
\newcommand{\tool}{\text{RVHyper}}

\newcommand{\eahyper}{\text{EAHyper}}

\newcommand{\ninfluences}{\not\leadsto}

\title{Efficient Monitoring of Hyperproperties using Prefix Trees
}
\author{Bernd Finkbeiner \and Christopher Hahn \and Marvin Stenger \and \\  Leander Tentrup}
\institute{Reactive Systems Group\\Saarland University\\\email{lastname@react.uni-saarland.de}}

\date{Received: date / Accepted: date}

\begin{document}

\maketitle
\sloppy

\begin{abstract}
	
	
  Hyperproperties, such as non-interference and observational determinism, relate multiple computation traces with each other and are thus not monitorable by tools that consider computations in isolation.
  We present the monitoring approach implemented in the latest version of $\tool$, a runtime verification tool for hyperproperties.
  The input to the tool are specifications given in the temporal logic $\hyperltl$, which extends linear-time temporal logic~(LTL) with trace quantifiers and trace variables.
  $\tool$ processes execution traces sequentially until a violation of the specification is detected.
  In this case, a counter example, in the form of a set of traces, is returned.
  $\tool$ employs a range of optimizations: a preprocessing analysis of the specification and a procedure that minimizes the traces that need to be stored during the monitoring process.
  In this article, we introduce a novel trace storage technique that arranges the traces in a tree-like structure to exploit partially equal traces.
  We evaluate $\tool$ on existing benchmarks on secure information-flow control, error correcting codes and symmetry in hardware designs.
  As an example application outside of security, we show how $\tool$ can be used to detect spurious dependencies in hardware designs.
  
\end{abstract}

{\let\thefootnote\relax\footnotetext{{This work was partially supported by the German Research Foundation (DFG) as part of the Collaborative Research Center ``Methods and Tools for Understanding and Controlling Privacy'' (CRC 1223) and the Collaborative Research Center ``Foundations of Perspicuous Software Systems'' (TRR 248, 389792660), and by the European Research Council (ERC) Grant OSARES (No. 683300).}}}

\section{Introduction}
\label{sec:intro}

\emph{Hyperproperties}~\cite{journals/jcs/ClarksonS10} are widely studied in (but not limited to) the context of secure information-flow control. They generalize trace properties in that they not only check the correctness of \emph{individual} computation traces in isolation, but relate \emph{multiple} computation traces to each other.
Examples include information-flow policies, such as observational determinism~\cite{journals/jcs/McLean92,conf/sp/Roscoe95,conf/csfw/ZdancewicM03}, (quantitative) noninterference~\cite{journals/jcs/McLean92,conf/fossacs/Smith09,conf/esorics/YasuokaT10} as well as symmetry~\cite{conf/cav/FinkbeinerRS15} and spurious dependencies in hardware designs~\cite{conf/tacas/FinkbeinerHST18}, error correcting codes~\cite{conf/cav/FinkbeinerRS15}, and anti-doping of automotive software~\cite{conf/esop/DArgenioBBFH17}.

In this article, we present the monitoring approach implemented in the latest version of $\tool$, an automata-based monitoring tool for hyperproperties~\cite{journals/fmsd/FinkbeinerH19}.
In dynamic verification of hyperproperties, efficient and light-weight monitoring techniques are instrumented in systems, which are usually far beyond the scope of static verification approaches. By doing so, countermeasures are enacted before, for example, irreparable information leaks happen.
A runtime verification tool for hyperproperties is in particular useful if the implementation of a security critical system is not available. Even without access to the source code, monitoring the observable execution traces still detects insecure information flow.
$\tool$ also supports the verification workflow by providing a method to test and develop specifications: Specifications can be checked on sample traces without the need for a complete model.
Based on the feedback of $\tool$, the specification can be refined until it matches the intended meaning.


The input of $\tool$ is given in the temporal logic $\hyperltl$~\cite{conf/post/ClarksonFKMRS14}, which expresses temporal hyperproperties by extending linear-time temporal logic with \emph{explicit} trace quantification (see~\cite{conf/lics/CoenenFHH19} for a recent study of hyperlogics).
HyperLTL has been used extensively to specify hyperproperties of practical interest (e.g~\cite{conf/cav/FinkbeinerRS15,journals/fmsd/FinkbeinerH19,conf/tacas/FinkbeinerHST18,conf/cav/FinkbeinerHT18,conf/cav/FinkbeinerHLST18,conf/esop/DArgenioBBFH17}).
For example, observational determinism is expressed as the following formula:
$$
\forall \pi. \forall \pi'. (o_\pi = o_{\pi'}) \LTLweakuntil (i_\pi \neq i_{\pi'})\enspace ,
$$
stating that every trace pair $\pi,\pi'$ has to agree on the output as long as it agrees on the inputs as well.
When detecting a violation, $\tool$ outputs a counter example, which is a set of traces that does not satisfy the input formula.

Efficient model checking, synthesis and satisfiability checking tools for HyperLTL already exist~\cite{conf/cav/FinkbeinerRS15,conf/cav/FinkbeinerHT18,conf/concur/FinkbeinerH16,conf/atva/FinkbeinerHH18,conf/cav/FinkbeinerHS17,conf/cav/FinkbeinerHLST18,conf/cav/CoenenFST19}. Implementing an efficient runtime verification tool for HyperLTL specifications is, despite recent theoretical progress~\cite{conf/csfw/AgrawalB16,conf/tacas/BrettSB17,conf/csfw/BonakdarpourF18,conf/isola/BonakdarpourSS18,conf/tacas/HahnST19,journals/fmsd/FinkbeinerH19,DBLP:conf/fm/StuckiSSB19,DBLP:conf/rv/Hahn19} difficult:
%
In principle, the monitor not only needs to process every observed trace, but must also \emph{store} every trace observed so far, so that future traces can be compared with the traces seen so far.

The previous version of $\tool$ tackles this challenging problem by implementing two optimizations~\cite{journals/fmsd/FinkbeinerH19,conf/tacas/FinkbeinerHST18}: a \emph{specification analysis} to detect exploitable properties of a hyperproperty, such as \emph{symmetry} and a \emph{trace analysis}, which detects all redundant traces that can be omitted during the monitoring process. A limitation of the trace analysis, which is based on a language inclusion check, is that only \emph{entire} traces can be analyzed and pruned.
For example, consider the traces $t_1 = \{a\}\{a\}\{\}$ and $t_2 = \{a\}\{\}\{a\}$ of length $3$ and the HyperLTL formula $\forall \pi. \forall \pi'. \G ( a_\pi \rightarrow \neg b_{\pi'} )$. Neither $t_1$ nor $t_2$ is dominated by the other trace, in the sense of the trace analysis, i.e., that one of the traces poses strictly less requirements on future traces~\cite{journals/fmsd/FinkbeinerH19}.
The traces, however, are equal on the first position.
This provides an opportunity for optimization, which our new approach exploits.
We introduce a novel trace storage technique (that also has massive impact on the running time), such that $\tool$ can also handle \emph{partially equal} traces by storing them in a tree structure.

We evaluate $\tool$ on existing benchmarks such as classical information-flow security by checking for violations of noninterference or monitoring error-resistant encoder. $\hyperltl$ is, however, not limited to security policies. As an example of such an application beyond security, we show how $\tool$ can be used to detect spurious dependencies in hardware designs.

\paragraph{Structure of this Article.}

The remainder of this article is structured as follows.
We begin by giving preliminaries on $\hyperltl$, its finite trace semantics and notation in Section~\ref{sec:prelims}. In Section~\ref{sec:rvhyper}, we present automata-based monitoring approach implemented in $\tool$, before discussing optimizations in Section~\ref{sec:opt} that make the monitoring feasible in practice. In Section~\ref{sec:eval}, we evaluate $\tool$ with a focus on the novel storage optimization technique using our tree data structure.

This is a revised and extended version of a paper that appeared at TACAS~2018~\cite{conf/tacas/FinkbeinerHST18}.
Our contribution and extension compared to~\cite{conf/tacas/FinkbeinerHST18} is the inclusion of a new trace storage optimization technique presented in Sec.~\ref{sec:trie} and an extended evaluation in Sec.~\ref{sec:eval}.

\paragraph{Related Work.}

The temporal logic $\hyperltl$ was introduced to model check security properties of reactive systems~\cite{conf/post/ClarksonFKMRS14,conf/cav/FinkbeinerRS15}.
For one of its predecessors, $\secltl$~\cite{conf/vmcai/DimitrovaFKRS12}, there has been a proposal for a white-box monitoring approach~\cite{conf/isola/DimitrovaFR12} based on alternating automata.
A recent survey on algorithms for monitoring hyperproperties is given in~\cite{DBLP:conf/rv/Hahn19}.
Agrawal and Bonakdarpour~\cite{conf/csfw/AgrawalB16} were the first to study the monitoring problem of HyperLTL for the sequential model. They give a syntactic characterization of monitorable $\hyperltl$ formulas. They present a first monitoring algorithm based on a progression logic expressing trace interdependencies and the composition of an LTL$_3$ monitor.
A first constraint-based approach has been outlined in~\cite{conf/tacas/BrettSB17}, which works for a subclass of HyperLTL specifications. The idea is to identify a set of propositions of interest and store corresponding constraints.
A constraint-based algorithm for the complete fragment of $\forall^2$ HyperLTL formulas has been proposed in~\cite{conf/tacas/HahnST19}. The algorithms rewrites a HyperLTL formula and an incoming event into a constraint composed of a plain LTL requirement as well as a HyperLTL requirement.
An constraint system is built incrementally: the HyperLTL part is encoded with variables, which will be incrementally defined with more incoming events of a trace.
Like with our monitoring algorithm, they do not have access to the implementation (black box), but in contrast to our work, they do not provide witnessing traces as a monitor verdict.

In~\cite{conf/csfw/BonakdarpourF18}, the authors study the complexity of monitoring hyperproperties. They show that the form and size of the input, as well as the formula have a significant impact on the feasibility of the monitoring process. They differentiate between several input forms and study their complexity: a set of linear traces, tree-shaped Kripke structures, and acyclic Kripke structures. For acyclic structures and alternation-free HyperLTL formulas, the problems complexity gets as low as NC.
In~\cite{conf/isola/BonakdarpourSS18,DBLP:conf/fm/StuckiSSB19}, the authors study where static analysis can be combined with runtime verification techniques to monitor HyperLTL formulas beyond the alternation-free fragment.

For certain information flow policies, like non-interference and some extensions, dynamic enforcement mechanisms have been proposed.
Techniques for the enforcement of information flow policies include tracking dependencies at the hardware level~\cite{conf/asplos/SuhLZD04}, language-based monitors~\cite{journals/jsac/SabelfeldM03,conf/csfw/AskarovS09,conf/pldi/AustinF10,conf/csfw/VanhoefGDPR14,conf/post/BichhawatRGH14}, and abstraction-based dependency tracking~\cite{conf/asian/GuernicBJS06,conf/essos/KovacsS12,conf/csfw/ChudnovKN14}.
Secure multi-execution~\cite{conf/sp/DevrieseP10} is a technique that can enforce non-interference by executing a program multiple times in different security levels.
To enforce non-interference, the inputs are replaced by default values whenever a program tries to read from a higher security level.

\section{Preliminaries}
\label{sec:prelims}
Let $\mathit{AP}$ be a finite set of \emph{atomic propositions} and let $\Sigma = 2^\mathit{AP}$ be the corresponding \emph{alphabet}.
An infinite \emph{trace} $t \in  \Sigma^\omega$ is an infinite sequence over the alphabet.
A subset $T \subseteq \Sigma^\omega$ is called a \emph{trace property}. A \emph{hyperproperty} $H \subseteq 2^{(\Sigma^\omega)}$ is a generalization of a trace property.
A finite trace $t \in \Sigma^+$ is a finite sequence over $\Sigma$.
In the case of finite traces, $|t|$ denotes the length of a trace.
We use the following notation to access and manipulate traces:
Let $t$ be a trace and $i$ be a natural number.
$t[i]$ denotes the $i$-th element of $t$.
Therefore, $t[0]$ represents the first element of the trace.
Let $j$ be natural number. If $j \geq i$ and $i\leq|t|$, then
$t[i,j]$ denotes the sequence $t[i] t[i+1] \cdots t[min(j,|t|-1)]$. Otherwise it denotes the empty trace $\epsilon$.
$t[i\rangle$ denotes the suffix of $t$ starting at position $i$.
For two finite traces $s$ and $t$, we denote their concatenation by $s \cdot t$.

\paragraph{HyperLTL Syntax.}\label{thltl:syntax}
HyperLTL~\cite{conf/post/ClarksonFKMRS14} extends LTL with trace variables and trace quantifiers.
Let $\mathcal{V}$ be a finite set of trace variables.
The syntax of HyperLTL is given by the grammar
\begin{alignat*}{3}
  \varphi  &{}\coloneqq ~ \forall \pi . \; \varphi \mid \exists \pi . \; \varphi \mid \psi  \\
  \psi &{}\coloneqq ~ a_{\pi} \mid \psi \wedge \psi \mid \neg \psi \mid \LTLnext \psi  \mid \psi \; \LTLuntil \;  \psi \enspace ,
\end{alignat*}
where $a \in \mathit{AP}$ is an atomic proposition and $\pi \in \mathcal{V}$ is a trace variable.
Atomic propositions are indexed by trace variables. The explicit trace quantification enables us to express properties like ``on all traces $\varphi$ must hold'', expressed by $\forall \pi \ldot \varphi$. Dually, we can express ``there exists a trace such that $\varphi$ holds'', expressed by $\exists \pi \ldot \varphi$.
We use the standard derived operators \emph{release} $\varphi \LTLrelease \psi \coloneqq \neg(\neg \varphi \LTLuntil \neg\psi)$, \emph{eventually} $\LTLdiamond \varphi \coloneqq \mathit{true} \LTLuntil \varphi$, \emph{globally} $\LTLsquare \varphi \coloneqq \neg \LTLdiamond \neg \varphi$, and \emph{weak until} $\varphi_1 \LTLweakuntil \varphi_2 \coloneqq (\varphi_1 \LTLuntil \varphi_2) \vee \LTLsquare \varphi_1$.
As we use the finite trace semantics, $\LTLnext \varphi$ denotes the \emph{strong} version of the next operator, i.e., if a trace ends before the satisfaction of $\varphi$ can be determined, the satisfaction relation, defined below, evaluates to false.
To enable duality in the finite trace setting, we additionally use the \emph{weak} next operator $\LTLweaknext \varphi$ which evaluates to true if a trace ends before the satisfaction of $\varphi$ can be determined and is defined as $\LTLweaknext \varphi \coloneqq \neg \LTLnext \neg \varphi$.
We call $\psi$ of a HyperLTL formula $\vec{Q}. \psi$, with an arbitrary quantifier prefix $\vec{Q}$, the \emph{body} of the formula. A HyperLTL formula $\vec{Q}. \psi$ is in the \emph{alternation-free fragment} if either $\vec{Q}$ consists solely of universal quantifiers or solely of existential quantifiers.
We also denote the respective alternation-free fragments as the $\forall^n$ fragment and the $\exists^n$ fragment, with $n$ being the number of quantifiers in the prefix.

\paragraph{Finite Trace Semantics.}
We recap the finite trace semantics for HyperLTL~\cite{conf/tacas/BrettSB17}, which is itself based on the finite trace semantics of $\ltl$~\cite{books/daglib/0080029}.
Let $\pathassignfin : \pathvars \rightarrow \Sigma^+$ be a partial function mapping trace variables to finite traces. We define $\epsilon[0]$ as the empty set.
$\pathassignfin[i\rangle$ denotes the trace assignment that is equal to $\pathassignfin(\pi)[i\rangle$ for all $\pi \in \dom(\pathassignfin)$.
By slight abuse of notation, we write $t \in \pathassignfin$ to access traces $t$ in the image of $\pathassignfin$.
The satisfaction of a HyperLTL formula $\varphi$ over a finite trace assignment $\pathassignfin$ and a set of finite traces $T$, denoted by $\pathassignfin \models_T \varphi$, is defined as follows:
\begin{equation*}
\begin{array}{ll}
    \pathassignfin \models_T a_\pi         \qquad & \text{if } a \in \pathassignfin(\pi)[0] \\
    \pathassignfin \models_T \neg \varphi              & \text{if } \pathassignfin \nmodels_T \varphi \\
    \pathassignfin \models_T \varphi \lor \psi         & \text{if } \pathassignfin \models_T \varphi \text{ or } \pathassignfin \models_T \psi \\
    \pathassignfin \models_T \LTLnext \varphi          & \text{if } \forall t \in \pathassignfin \ldot |t|>1 \text{ and } \pathassignfin[1\rangle \models_T \varphi \\
    \pathassignfin \models_T \varphi\LTLuntil\psi      & \text{if } \exists i < \min\nolimits_{t \in \pathassignfin} |t| \ldot \pathassignfin[i\rangle \models_T \psi \\
    &~\hspace{1ex} \land \forall j < i \ldot \pathassignfin[j\rangle \models_T \varphi \\
    \pathassignfin \models_T \exists \pi \ldot \varphi & \text{if } \exists t \in T \text{ such that } \pathassignfin[\pi \mapsto t] \models_T \varphi \\
    \pathassignfin \models_T \forall \pi \ldot \varphi & \text{if } \forall t \in T \text{ holds that } \pathassignfin[\pi \mapsto t] \models_T \varphi
\end{array}
\end{equation*}
%
The hyperproperty represented by a $\hyperltl$ formula $\varphi$, denoted by $\hlang(\varphi)$, is the set $\set{ T \subseteq \Sigma^\omega \mid T \models \varphi }$.
%

\section{Runtime Verification of Hyperproperties with $\tool$}
\label{sec:rvhyper}
In this section, we present an overview over $\tool$, before describing the implementation setup, present the monitoring algorithm, and discuss our optimization techniques.

The input of $\tool$ is given as a universally quantified HyperLTL formula and, in addition, the observed behavior of the system under consideration.
The observed behavior is represented as a trace set $T$, where each $t \in T$ represents a previously observed execution of the system to monitor.
$\tool$ can therefore detect violations of every monitorable $k$-safety hyperproperty (see~\cite{journals/fmsd/FinkbeinerH19} for an extensive study of monitorability of hyperproperties).
If $\tool$ detects that the system violates the hyperproperty, it outputs a counter example, i.e, a $k$-ary tuple of traces, where $k$ is the number of quantifiers in the HyperLTL formula.

\subsection{Implementation Details}
$\tool$\footnote{The implementation is available at \url{https://react.uni-saarland.de/tools/rvhyper/}.} is written in C\nolinebreak[4]\hspace{-.05em}\raisebox{.4ex}{\relsize{-3}{\textbf{++}}}.
We use \emph{spot}~\cite{conf/atva/Duret-LutzLFMRX16} for building the deterministic monitor automata and the \emph{Buddy} BDD library for handling symbolic constraints.
We use the $\hyperltl$ satisfiability solver $\eahyper$~\cite{conf/cav/FinkbeinerHS17,conf/concur/FinkbeinerH16} to determine whether the input formula is reflexive, symmetric, or transitive.
Depending on those results, we omit redundant tuples in the monitoring algorithm.

\subsection{Online Monitoring Algorithm}

For the online algorithm, we use standard techniques for building LTL monitoring automata and use this to instantiate this monitor by the traces as specified by the $\hyperltl$ formula.
Let $\ap$ be a set of atomic propositions and $\pathvars = \{\pi_1, \ldots, \pi_n\}$ a set of trace variables. 
A deterministic monitor template $\monitor = (\Sigma, Q, \delta, q_0, F)$ is a tuple of
a finite alphabet $\Sigma = 2^{(\ap \times \pathvars)}$,
a non-empty set of states $Q$,
a partial transition function $\delta: Q \times \Sigma \hookrightarrow Q$,
a designated initial state $q_0 \in Q$, and
a set of accepting states $F \subseteq Q$.
The instantiated automaton runs in parallel over traces in $(2^\ap)^*$, thus we define a run with respect to a $n$-ary tuple $N \in ((2^\ap)^*)^n$ of finite traces.
A run of $N$ is a sequence of states $q_0 q_1 \cdots q_m \in Q^*$, where $m$ is the length of the smallest trace in $N$, starting in the initial state $q_0$ such that for all $i$ with $0 \leq i < m$ it holds that 
\begin{equation*}
\delta\left(q_i, \bigcup_{j=1}^n \bigcup_{a \in N(j)(i)} \set{(a,\pi_j)} \right) = q_{i+1} \enspace.
\end{equation*}
A tuple $N$ is accepted if there is a run on $\monitor$ that ends in an accepting state.
For LTL, such a deterministic monitor can be constructed in doubly-exponential time in the size of the formula~\cite{conf/cav/dAmorimR05,journals/fmsd/TabakovRV12}.

\begin{figure}[t]
	\centering
 \begin{tikzpicture}[auto,->,>=stealth',shorten >=1pt,thick,transform shape, scale=0.9]
 \node[state,initial,initial text=,accepting] (init) {$q_0$};
 \node[state,below left=1.3 and 1.5 of init,accepting] (accepting) {$q_\top$};
 \node[state,black!50,below right=1.3 and 1.5 of init] (rejecting) {$q_\bot$};
 
 \draw (init) edge[loop right] node[swap] {$o_\pi = o_{\pi'} \land i_\pi = i_{\pi'}$} ()
       (init) edge node {$i_\pi \neq i_{\pi'}$} (accepting)
       (init) edge[black!50] node[near end] {$o_\pi \neq o_{\pi'} \land i_\pi = i_{\pi'}$} (rejecting)
       (accepting) edge[loop right] node {$\top$} (accepting)
       (rejecting) edge[loop right,black!50] node {$\top$} (rejecting)
 ;
 \end{tikzpicture}
	\caption{Visualization of a monitor template corresponding to the formula given in Example~\ref{ex:obsdet-monitor}. We use a symbolic representation of the transition function $\delta$. We depict a sink rejecting state $q_\bot$.}
	\label{fig:monitor-template}
\end{figure}
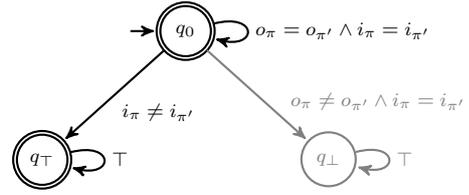

\begin{example}  \label{ex:obsdet-monitor}
  As an example formula, consider again the observational determinism formula introduced in the introduction:
  \begin{equation*}
    \forall \pi. \forall \pi'. (o_\pi = o_{\pi'}) \LTLweakuntil (i_\pi \neq i_{\pi'})\enspace ,
  \end{equation*}
  The corresponding monitor template is depicted in Fig.~\ref{fig:monitor-template}.
\end{example}

The algorithm for monitoring $\hyperltl$ formulas when traces are given sequentially to the monitor is presented as Algorithm~\ref{alg:algorithms-sequential}.
After building the deterministic monitoring automaton $\monitor_\varphi$, the algorithm accepts new traces and afterwards proceeds with the pace of the incoming stream.
We have a variable $S$ that maps tuples of traces to states of the deterministic monitor.
Whenever the current trace $t$ progresses, we progress every tuple $(t_1,\ldots,t_n)$ that contains $t$ with one of the following outcomes:
\begin{enumerate}
  \item One of the traces $t_1,\ldots,t_n$ may have ended, thus, we check if the monitor is in an accepting state and report a violation if this is not the case.
  \item There is a successor state in the monitor, thus we update $S$.
  \item There is no successor state, hence, we report a violation.
\end{enumerate}
When a new trace $t$ starts, only new tuples are considered for $S$, that are tuples $\vec{t} \in (T \cup \set{t})^n$ containing the new trace $t$.
			\begin{algorithm*}[t]
				\SetKwInOut{Input}{input}
				\SetKwInOut{Output}{output}
				\SetAlgoLined
				\Input{$\forall^n$ HyperLTL formula $\varphi$}
				\Output{satisfied or $n$-ary tuple witnessing violation}
				\BlankLine
				$\monitor_\varphi = (\Sigma_\pathvars, Q, q_0, \delta, F) =$ \texttt{build\_template($\varphi$)}\;
				$T \gets \emptyset$\;
				$S: T^n \rightarrow Q$ \quad // initially empty\;
				$t$  \quad // container for the subsequently incoming event traces
				\BlankLine
				\While{there is a new event trace}{
                  $t \gets \epsilon$ \quad // initialize empty trace\;
                  \For(init $S$ for every new tuple $\vec{t}$){$\vec{t} \in ((T \cup \set{t})^n \setminus T^n)$}{
    				$S(\vec{t}) \gets q_0$\;
    			  }
                  \While{$p \in \Sigma$ is a new input event}{
						$t' \gets t~p$ \quad // append $p$ to $t$\;
                        \For(// progress every state in $S$) {$( (t_1,\ldots,t_n), q) \in S$ where $t \in \{t_1,\ldots,t_n\}$}{
                            \uIf(// some trace ended){$\exists t^* \in \{t_1,\ldots,t_n\} \ldot \card{t^*} < \card{t}$}{
                             \eIf {$q \in F$} {
                              remove $(t_1,\ldots,t_n)$ from $S$\;
                              \textbf{continue}\;
                             } { \Return violation and witnessing tuple $(t_1,\ldots,t_n)$\;}
                          }
                          $\vec{t} \gets (t_1,\ldots,t_n)$, where $t_i$ is replaced with $t'$, for each $t_i = t$\;
                          \uIf {$\delta(q, \bigcup_{i=1}^n \bigcup_{a \in \vec{t}(i)[\card{t'}-1]} \set{ (a, \pi_i) }) = q'$} {
        				     $S(\vec{t}) \gets q'$\;
        				  }
        				  \Else{
        				    \Return violation and witnessing tuple $\vec{t}$\;
        				  }
        				}
						$t \gets t'$ \quad // re-assign $t$\;
				  } 
				  $T \gets T \cup \set{t}$\; \quad //add $t$ to already seen traces
				}
				\Return satisfied\;
				\BlankLine
	\caption{Online monitoring algorithm for $\forall^n \hyperltl$: The algorithm subsequently reads event traces and monitors them against already seen traces. Already seen traces are stored in $T$ before continuing with a new event trace.}
	\label{alg:algorithms-sequential}
			\end{algorithm*}

\begin{example} \label{ex:obsdet-algorithm}
  We continue \autoref{ex:obsdet-monitor} by showing how the algorithm progresses on the given formula.
  Assume for the sake of readability that we have a single input proposition $i$ and a single output proposition $o$.
  Furthermore, assume that we have already seen the trace $t_0 = \set{i}\set{i,o}\set{o}$, that is, $T = \set{t_0}$ and $S(t_0,t_0) = q_0$.
  We now show how the algorithm continues with a fresh trace $t_1$.
  In lines 6--8 we add the pairs $(t_0,t_1)$, $(t_0,t_1)$, and $(t_1,t_1)$ with the initial state $q_0$ to $S$.
  Let $p = \set{i}$ be the first input proposition, thus, $t_1 = \set{i}$.
  Since $t_0[0] = t_1[0]$, the monitor remains in $q_0$ for every tuple.
  Let $p = \set{i}$ be the next input proposition, thus, $t_1 = \set{i}\set{i}$.
  Consider the tuple $(t_0,t_1)$.
  As $t_0[1]$ and $t_1[1]$ are equal with respect to $i$ but differ in $o$, the monitor progresses to the rejecting state $q_\bot$ and the algorithm terminates by reporting the violation.
  If, the previous input proposition is $p = \set{}$, both tuples, $(t_0,t_1)$ and $(t_1,t_0)$ would progress to the accepting sink state $q_\top$ as the input proposition is different to $t_0[1]$.
  Assume two more inputs, e.g., $t_1 = \set{i}\set{i}\set{}\set{}$, then the pairs $(t_0,t_0)$, $(t_0,t_1)$, and $(t_1,t_0)$ are removed from $S$ as $t_1$ is strict longer than $t_0$.
\end{example}

\section{Optimizations}
\label{sec:opt}
\vspace{-1ex}
In this section, we present three optimizations implemented in $\tool$, which, as we will see in the evaluation section, are necessary to make the automata-based monitoring approach feasible in practice. We begin by explaining a \emph{specification analysis}, which is a preprocessing step that exploits properties of the specification to reduce the algorithmic workload. In the subsequent section, we show how $\tool$ tackles the problem of potentially unbounded memory consumption: We recap the \emph{trace analysis}, which was so far the only storage optimization implemented in $\tool$. We then provide a novel storage optimization technique based on prefix-trees, so-called \emph{tries}, to exploit partial equality of the given traces.
\vspace{-1ex}
\subsection{Specification Analysis}
\vspace{-1ex}
In the example execution in \autoref{ex:obsdet-algorithm} we have seen that the algorithm had to do more work than necessary to monitor observational determinism.
For example, a tuple $(t,t)$ for some trace $t$ cannot violate observational determinism as the traces are equal.
Further, from the pairs $(t,t')$ and $(t',t)$ for some traces $t$ and $t'$, we only need to check one of them as a violation in one of them implies the violation in the other pair.
We can automatically check for such conditions and, thus, omit unnecessary work.

$\tool$ implements the specification analysis with the HyperLTL satisfiability solver EAHyper~\cite{conf/concur/FinkbeinerH16,conf/cav/FinkbeinerHS17}. The \emph{specification analysis} is a preprocessing step that analyzes the HyperLTL formula under consideration.
EAHyper can detect whether a formula is (1) \emph{symmetric}, i.e., we halve the number of instantiated monitors, (2) \emph{transitive}, i.e, we reduce the number of instantiated monitors to two, or (3) \emph{reflexive}, i.e., we can omit the self comparison of traces.

\begin{definition}\cite{journals/fmsd/FinkbeinerH19} \label{def:symmetry}
	Let $\psi$ be the quantifier-free part of some $\hyperltl$ formula $\varphi$ over trace variables $\pathvars$.
	We say $\varphi$ is invariant under trace variable permutation $\sigma : \pathvars \to \pathvars$, if for any set of traces $T \subseteq \Sigma^{\omega}$ and any assignment $\pathassign : \pathvars \to T$, $(\emptyset, \pathassign, 0) \models \psi \Leftrightarrow (\emptyset, \pathassign \circ \sigma, 0) \models \psi$.
	We say $\varphi$ is symmetric, if it is invariant under every trace variable permutation in $\pathvars$.
\end{definition}
Symmetry is particularly interesting since many information flow policies satisfy this property. Consider, for example, observational determinism: 
$
\forall \pi\ldot \forall \pi'\ldot (o_\pi = o_{\pi'}) \W (i_\pi \neq i_{\pi'}).
$ 
$\tool$ detects symmetry by translating this formula to a formula that is unsatisfiable if there exists no set of traces for which every trace pair violates the symmetry condition:
$
\exists \pi\ldot \exists \pi'\ldot \big((o_\pi = o_{\pi'}) \W (i_\pi \neq i_{\pi'})\big) \nleftrightarrow \big((o_\pi = o_{\pi'}) \W (i_\pi \neq i_{\pi'})\big)
$.
If the resulting formula turns out to be unsatisfiable, $\tool$ omits the symmetric instantiations of the monitor automaton.
\begin{definition}\cite{journals/fmsd/FinkbeinerH19} \label{def:transitivity}
	Let $\psi$ be the quantifier-free part of some $\hyperltl$ formula $\varphi$ over trace variables $\set{\pi_1, \pi_2}$.
	Let $T = \set{t_1, t_2, t_3} \in \Sigma^{\omega}$ be a three-elemented set of traces.
	We define the assignment $\pathassign_{i,j} : \pathvars \to \Sigma^{\omega}$ by $\pathassign_{i,j} \coloneqq \{\pi_1 \mapsto t_i, \pi_2 \mapsto t_j\}$.
	We say $\varphi$ is transitive, if for all three-elemented sets $T$ it holds that $(\emptyset, \pathassign_{1,2}, 0) \models \psi \wedge (\emptyset, \pathassign_{2,3}, 0) \models \psi \rightarrow (\emptyset, \pathassign_{1,3}, 0) \models \psi$.
\end{definition}
While symmetric HyperLTL formulas allow us to prune half of the monitor instances, transitivity of a HyperLTL formula has an even larger impact on the required memory. Equality, i.e, $\forall \pi. \forall \pi' \ldot \G (a_\pi \leftrightarrow a_{\pi'})$, for example, is transitive and symmetric and allows us to reduce the number of monitor instances to one, since we can check equality against any reference trace.
\begin{definition}\cite{journals/fmsd/FinkbeinerH19} \label{def:reflexivity}
	Let $\psi$ be the quantifier-free part of some $\hyperltl$ formula $\varphi$ over trace variables $\pathvars$.
	We say $\varphi$ is reflexive, if for any trace $t \in \Sigma^{\omega}$ and the corresponding assignment $\pathassign : \pathvars \to \{t\}$, $(\emptyset, \pathassign, 0) \models \psi$.
\end{definition}
Lastly, if a formula is reflexive, $\tool$ omits the composition of a trace with itself during the monitoring process.
For example, equality and observational determinism have reflexive HyperLTL formulas.

\begin{example}
  Consider again the observational determinism formula from \autoref{ex:obsdet-monitor}.
  We have seen that this formula is both, reflexive and symmetric, thus, we can omit those instances in the algorithm.  
\end{example}

\subsection{Optimizing Trace Storage}
\label{sec:trie}

The main obstacle in monitoring hyperproperties is the potentially unbounded space consumption.
Previously, RVHyper employed a \emph{trace analysis} technique to detect redundant traces, with respect to a given HyperLTL formula, i.e., traces that can be safely discarded without losing any information and without losing the ability to return a counter example.

\begin{definition}~\cite{journals/fmsd/FinkbeinerH19}
	Given a HyperLTL formula $\varphi$, a trace set $T$ and an arbitrary $t \in \traces$, we say that $t$ is $(T,\varphi)$-redundant if $T$ is a model of $\varphi$ if and only if $T \cup \set{t}$ is a model of $\varphi$ as well, formally
	\begin{equation*}
	\forall T' \supseteq T \ldot T' \in \hlang(\varphi) \Leftrightarrow T' \cup \{t\} \in \hlang(\varphi) \enspace.
	\end{equation*}
\end{definition}

\begin{definition}~\cite{journals/fmsd/FinkbeinerH19}
	Given $t,t' \in \Sigma^\omega$, we say $t$ \emph{dominates} $t'$ with respect to $\varphi$ (or simply $t$ \emph{dominates} $t'$ if it is clear from the context) if $t'$ is ($\{t\},\varphi$)-redundant.
\end{definition}

\begin{example}
  For observational determinism, a trace $t$ is dominated by a trace $t'$ if $\card{t} < \card{t'}$ and both traces agree on the input propositions.
\end{example}

This is efficiently implemented in $\tool$ (cf. Algorithm~\ref{alg:traceopti}) and is guaranteed to catch all redundant traces.
In our experiments~\cite{journals/fmsd/FinkbeinerH19,conf/tacas/FinkbeinerHST18}, we made the observation that traces often share the same prefixes, leading to a lot of redundant monitor automaton instantiations, repetitive computations and duplicated information when those traces get stored.

The trace analysis, as it is based on a language inclusion check of the entire traces, cannot handle \emph{partial redundancy}, for example, in the case that traces have redundant prefix requirements.
This leaves room for optimization, which we address by implementing a \emph{trie} data structure for managing the storage of incoming traces.

\begin{algorithm}
	\SetKwInOut{Input}{input}
	\SetKwInOut{Output}{output}
	\SetAlgoLined
	\Input{$\hyperltl$ formula $\varphi$, redundancy free trace set $T$, fresh trace $t$}
	\Output{redundancy free set of traces $T_\mathit{min} \subseteq T \cup \set{t}$}
	$\monitor_\varphi =$ \texttt{build\_template($\varphi$)}
	\BlankLine
	\ForEach{$t' \in T$}{
		\If{ $t'$ dominates $t$ }
		{
			\Return $T$\;
		}
	}
	\ForEach{$t' \in T$}{
		\If{$t$ dominates $t'$}
		{
			$T \coloneqq T \setminus \set{t'}$\;
		}
	}
	\Return $T \cup \set{t}$\;
	\caption{Trace analysis algorithm to minimize trace storage.}
	\label{alg:traceopti}
\end{algorithm}

Tries, also known as prefix trees are a tree data structure, which can represent a set of words over an alphabet in a compact manner.
The root of a trie is identified with the empty word $\epsilon$, additionally each node can have several child nodes, each of which corresponds to a unique letter getting appended to the representing word of the parent node. So the set of words of a trie is identified with the set of words the leaf nodes represent.

\begin{definition}
A trie is a four tuple $(\Sigma, \Tau, \longrightarrow, \tau_0)$ consisting of
\begin{itemize}
    \item a finite alphabet $\Sigma$,
    \item a non-empty set of states $\Tau$,
    \item a transition function $\longrightarrow: \Tau \times \Sigma \rightarrow \Tau$,
    \item and a designated initial state $\tau_0 \in \Tau$ called the root.
    \end{itemize}
\end{definition}
Instead of $((\tau,a),\tau') \in \longrightarrow$ we will write $\tau \overset{a}{\longrightarrow} \tau'$ in the following.
For a trie to be of valid form we restrict $\longrightarrow$ such that, $\forall \tau,\tau' \in \Tau. |\{\tau \overset{a}{\longrightarrow} \tau'| a \in \Sigma\}| \leq 1$.


In our case the alphabet would be the set of propositions used in the specification, and the word built by the trie represents the traces. 
Instead of storing each trace individually, we store all of them in one trie structure, branching only in case of deviation. This means equal prefixes only have to be stored once.
Besides the obvious benefits for memory, we also can make use of the maintained trie data structure to improve the runtime of our monitoring algorithms. As traces with same prefixes end up corresponding to the same path in the trie, we only have to instantiate the monitor automaton as much as the trie contains branches.
\begin{example}
	Consider the following traces of length $6$ over the alphabet $2^{\{i,o\}}$:
	\begin{itemize}
		\item
		$t_1: \{\set{i},\set{i,o},\set{i},\set{i},\set{i},\set{i,o} \}$
		\item
		$t_2: \{\set{i}, \set{i,o}, \set{i}, \set{i} , \set{i}, \set{i}\}$
		\item
		$t_3: \{\set{i}, \set{i}, \set{i}, \set{i} , \set{i}, \set{i}\}$
		\item
		$t_4: \{\set{i}, \set{i}, \set{}, \set{} , \set{}, \set{}\}$
	\end{itemize}
After processing the traces sequentially, the resulting trie looks as follows:

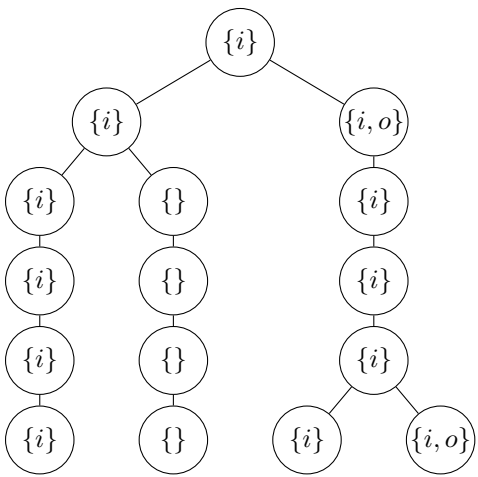
\vspace{1ex}

\begin{tikzpicture}[sibling distance=7em, level distance = 30pt,
every node/.style = {shape=circle,
	draw, align=center, minimum size = 0.9cm, inner sep = 0pt},
level 1/.style = {sibling distance = 10em},
level 2/.style = {sibling distance = 5em},
level 5/.style = {sibling distance = 5em}]
\node {$\{i\}$}
child { node {$\{i\}$}
	child { node {$\{i\}$}
		child { node {$\{i\}$}
			child { node {$\{i\}$}
				child { node {$\{i\}$} } } } } 
	child { node {$\{ \}$}  
		child { node {$\{ \}$} 
			child { node {$\{ \}$}
				child { node {$\{ \}$} } } } } }
child { node {$\{i,o\}$}
	child { node {$\{i\}$}
		child { node {$\{i\}$}
			child { node {$\{i\}$}
				child { node {$\{i\}$} }
				child { node {$\{i,o\}$} } } } } };
\end{tikzpicture}
\end{example}

\subsection{Trie-based Monitoring Algorithm}

We depict a trie-based offline monitoring algorithm in Fig.~\ref{alg:trie-offline}.
For the sake of readability, we assume that there are as many traces as universal quantifiers, that we progress through all traces in parallel, and that all traces have the same length.
This is merely a simplification in the presentation, one can build the trie in a sequential fashion for online monitoring by a slight modification of the presented algorithm.

Without using tries, our monitoring algorithm was based on instantiating the deterministic monitor template $\monitor_\varphi$ with tuples of traces. Now we instantiate $\monitor_\varphi$ with tuples of tries. Initially we only have to create the single instance having the the root of our trie.

	\begin{algorithm*}[t]
		\SetKwInOut{Input}{input}
		\SetKwInOut{Output}{output}
		\SetAlgoLined
		\Input{$\forall^n$ HyperLTL formula $\varphi$}
		\Output{satisfied or $n$-ary tuple witnessing violation}
		\BlankLine
		$\monitor_\varphi = (\Sigma_\pathvars, Q, q_0, \delta, F) =$ \texttt{build\_template($\varphi$)}\\
		$S: \Tau^n \rightarrow Q$\\
		$\tau_0 \coloneqq $\texttt{new\_trie()}\\
		$\mathbf{i} \coloneqq (\tau_0,\ldots,\tau_0) \in \Tau^n$\\
		$I \coloneqq \{\mathbf{i}\}$  \quad // set of not-yet terminated branches\\
		\BlankLine
		\While{$\mathbf{p} \leftarrow$ new event (in $\Sigma^n$)}{
			\For{$1 \leq j \leq n$}{
				$\mathbf{i}(j) \leftarrow $\texttt{add\_child($\mathbf{i}(j)$, $\mathbf{p}(j)$)} \quad // add child with value $\vec{p}(j)$ to $\vec{i}(j)$ if needed \\
			}
			// update set of active branches\\
			$I \leftarrow \bigcup_{\mathbf{i} \in I} \set{ (i'_1,\ldots, i'_n) \mid \mathbf{i}(j) \overset{a}{\longrightarrow} i'_j, a \in \Sigma, 1 \leq j \leq n }$\\
			\ForEach{$\mathbf{i} \in I$}{
				progress every state in $S$ according to $\delta$\;
				\If{violation in $\monitor_\varphi$}{
				    // return sequence from root to $\vec{i}$\\
					$t \gets ($\texttt{rooted\_sequence($\vec{i}(1)$)}$,\ldots,$\texttt{rooted\_sequence($\vec{i}(n)$)}$)$\\
					\Return witnessing tuple $t^n$\\
				}
			}
		}
		\Return satisfied if $\forall \vec{i} \in I \ldots S(\vec{i}) \in F$ else violation\\
		\BlankLine
	\caption{Offline algorithm using trie data structure.}
	\label{alg:trie-offline}
\end{algorithm*}

The trie-based algorithm has much in common with its previously discussed trace-based counterpart. Initially, we have to build the deterministic monitor automaton $\monitor_\varphi = (\Sigma_\pathvars, Q, q_0, \delta, F)$.
We instantiate the monitor with a fresh trie root $\tau_0$.
A mapping from trie instantiations to a state in $\monitor_\varphi$ $S: \Tau^n \rightarrow Q$, stores the current state of every active branch of the trie, stored in the set $I$.
For each of the incoming traces, we provide an entry in a tuple of tries $\mathbf{\tau}$, each entry gets initialized to $\tau_0$.
During the run of our algorithm these entries are updated such that they always correspond to the word built by the traces up to this point.
For as long as there are traces left, which have not yet ended, and we have not yet detected a violation, we will proceed updating the entries in $\vec{i}$ as follows.
Having entry $\tau$ and the correspond trace sequence proceeds with $a$, if $\exists \tau' \in \Tau. \tau \overset{a}{\longrightarrow} \tau'$, we update the entry to $\tau'$ otherwise we create such a child node of $\tau$ (\texttt{add\_child} in line 8).
Creating a new node in the trie always occurs when the prefix of the incoming trace starts to differ from already seen prefixes.
After having moved one step in our traces sequences, we have to reflect this step in our trie structure, in order for the trie-instantiated automata to correctly monitor the new propositions.
As a trie node can branch to multiple child nodes, each monitor instantiation are replaced by the set of instantiations, where all possible child combinations of the different assigned tries are existent (update of $I$ in line 11).
Afterwards, we update $S$ in the same way as in Algorithm~\ref{alg:traceopti}, thus, we omit algorithmic details here.
If a violation is detected here, that is there is no transition in the monitor corresponding to $\vec{i}$, we will return the corresponding counter example as a tuple of traces, as those can get reconstructed by stepping upwards in the tries of $\vec{i}$.
If the traces end, we check if every open branch $\vec{i} \in I$ is in an accepting state.


\section{Evaluation}
\label{sec:eval}
In the following, we evaluate the new version of $\tool$, especially the novel trace storage optimization. We use several benchmarks: an encoder that guarantees a Hamming-distance of $2$, violations of noninterference on randomly generated traces, and a symmetry property on an implementation of the Bakery protocol.
As an example how $\tool$ can be used outside security runtime verification, we give a case study on detecting spurious dependencies in hardware designs.

\subsection{Error Correcting Codes}
\begin{figure*}[t]
	\centering
	\scalebox{0.75}{
	\begin{tikzpicture}
	\begin{axis}[width=0.55\textwidth,mark size=1.3pt,ymode=log,ymax=400000,xmode=log,xmin=1,xmax=5000,no markers,thick,xlabel={probability for input bit flip $\times 10^{-4}$},ylabel={runtime in msec.},
	legend entries={naive,spec analysis,trace analysis,spec \& trace analysis,tries,spec analysis \& tries},
	legend style={
		at={(-0.2,1)},
		anchor=north east}]
	]]
	\addplot+[red,solid] table {mux30naive.tex};
	\addplot+[blue,dashed,very thick] table {mux30sa.tex};
	\addplot+[orange,dotted,very thick] table {mux30ta.tex};
	\addplot+[green,dashdotted,very thick] table {mux30both.tex};
	\addplot+[yellow,dashdotted,very thick] table {mux30trienaive.tex};
	\addplot+[violet,dotted,very thick] table {mux30trie.tex};
	\end{axis}
	\end{tikzpicture}\begin{tikzpicture}
	\begin{axis}[width=0.55\textwidth,mark size=1.3pt,ymin=0,ymax=56296,xmin=0,xmax=1000,no markers,thick,xlabel={\# of instances},ylabel={runtime in msec.},
	]
	]]
	\addplot+[red,solid] table {encnaive.tex};
	\addplot+[blue,dashed,very thick] table {encsa.tex};
	\addplot+[orange,dotted,very thick] table {encta.tex};
	\addplot+[green,dashdotted,very thick] table {encboth.tex};
	\addplot+[yellow,dashdotted,very thick] table {enctrienaive.tex};
	\addplot+[violet,dotted,very thick] table {enctrie.tex};
	\end{axis}
	\end{tikzpicture}
}
	\caption{Left: Monitoring of black box circuits (mux example). Right: Hamming-distance preserving encoder; runtime comparison of naive monitoring approach with different optimizations and a combination thereof.}
	\label{fig:runtime-comparison-prob}
\end{figure*}
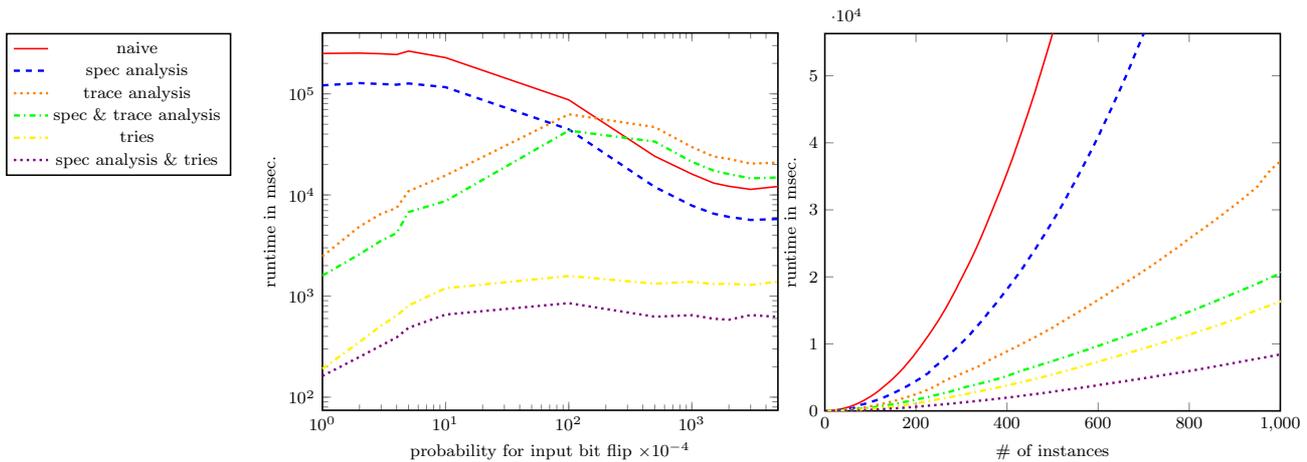

We monitored whether an encoder preserves a Hamming-distance of $2$. We randomly built traces of length $50$. In each position of the trace, the corresponding bit had a 1\% chance to be flipped. The specification can be encoded as the following $\hyperltl$ formula~\cite{conf/cav/FinkbeinerRS15}:
\begin{align*}
\forall \pi \pi'. &(\LTLdiamond (i_\pi \nleftrightarrow i_{\pi'}) \rightarrow ((o_\pi \leftrightarrow o_{\pi'}) \\ &\LTLuntil ((o_{\pi}  \nleftrightarrow o_{\pi'})  \wedge \LTLnext ((o_\pi \leftrightarrow o_{\pi'}) \LTLuntil (o_{\pi} \nleftrightarrow o_{\pi'}))))).
\end{align*}
The right plot of Figure~\ref{fig:runtime-comparison-prob} shows the results of our experiments. We compared the naive monitoring approach to different combinations of $\tool$'s optimizations. The specification analysis returns in under one second with the result that the formula is symmetric and reflexive. Hence, as expected, this preprocessing step has a major impact on the running time of the monitoring process as more than half of the, in general necessary, monitor instantiations can be omitted. A combination of the specification and trace analysis performs nearly equally well as naively storing the traces in our trie data structure. Combining the trie data structure with the specification analysis performs best and results in a tremendous speed-up compared to the naive approach.

\subsection{Checking Noninterference}

\begin{table*}
	\caption{Non-Interference Benchmark: Monitored $2000$ traces of length $50$ with an increasing input size.}
	\label{tbl:rvhyper-ni}
	\centering
	\begin{tabular}{rlllrrrr}
		\hline \noalign{\smallskip}
		instance & \multicolumn{3}{c}{only spec analysis} & \multicolumn{4}{c}{tries+spec analysis} \\
		& \#\,instances & \#\,transitions & time & \#\,instances & \#\,transitions & \#\,trie nodes & time  \\
		\noalign{\smallskip}\hline\noalign{\smallskip}
		8-bit & 1999000 & 4312932 & 14807ms &   2 &  26734 & 11262 &  226ms \\
		16-bit & 1999000 & 2772001 & 11166ms &   4 &  34365 & 87258 &  285ms \\
		24-bit & 1999000 & 2401723 & 11330ms &   8 &  45757 & 93353 &  416ms \\
		32-bit & 1999000 & 2236529 & 13814ms &  16 &  68364 & 95237 &  636ms \\
		40-bit & 1999000 & 2148818 & 15353ms &  32 & 103315 & 96273 & 1033ms \\
		48-bit & 1999000 & 2102689 & 18769ms &  64 & 163888 & 96941 & 1994ms \\
		56-bit & 1999000 & 2074460 & 22310ms & 128 & 268094 & 97506 & 3580ms \\
		64-bit & 1999000 & 2063497 & 32617ms & 248 & 434705 & 97831 & 7561ms \\
		\hline
	\end{tabular}
\end{table*}

Non-interference~\cite{journals/jcs/McLean92} is an important information flow policy demanding that an observer of a system cannot infer any high security input of a system by observing only low security input and output.
Formally, we specify that all low security outputs $\vec{o}^{low}$ have to be equal on all system executions as long as the low security inputs $\vec{i}^{low}$ of those executions are the same:
$\forall \pi, \pi' \ldot ({\vec{o}^{low}_\pi} \leftrightarrow {\vec{o}^{low}_{\pi'}}) \LTLweakuntil ({\vec{i}^{low}_\pi} \nleftrightarrow {\vec{i}^{low}_{\pi'}}).$
This class of benchmarks has previously been used to evaluated RVHyper~\cite{conf/tacas/FinkbeinerHST18}.
We repeated the experiments, to show that using the trie data structure is a valid optimization.
The results are depicted in Table~\ref{tbl:rvhyper-ni}.
We chose a trace length of $50$ and monitored non-interference on $2000$ randomly generated traces, where we distinguish between an input range of $8$ to $64$ bits.
The results show, that the trie optimization has an enormous impact compared to a naive approach that solely relies on the specification analysis. As expected, the difference in runtime is especially high on experiments where traces collapse heavily in the trie data structure, i.e., producing almost no instances that must be considered during the monitoring process.

\subsection{Symmetry in Mutual Exclusion Protocols}
\begin{figure*}[t]
	\resizebox{\textwidth}{!}{
		\input{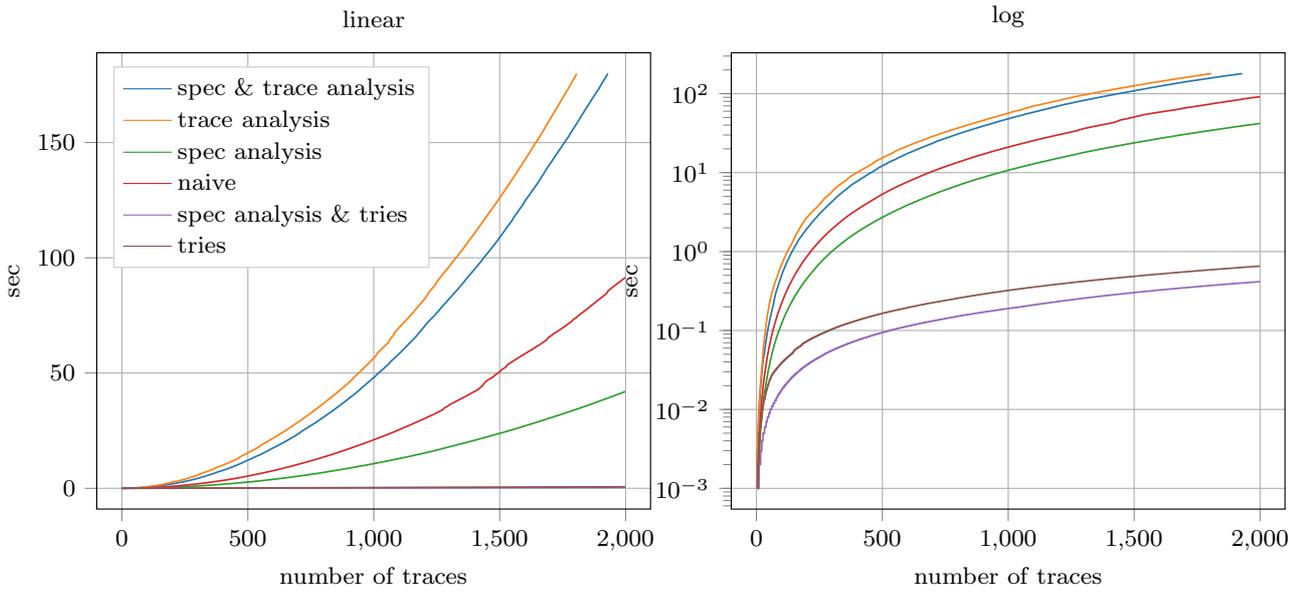}
	}
	\caption{Experiment of monitoring symmetry on an implementation of the bakery protocol.}
	\label{fig:bakery}
\end{figure*}

In this benchmark (introduced as a case study in~\cite{conf/cav/FinkbeinerRS15}), we monitor whether a Verilog implementation of the bakery protocol~\cite{journals/cacm/Lamport74a} from the VIS verification benchmark satisfies a symmetry property. Symmetry violations indicate that certain clients are privileged.
The Bakery protocol is a classical protocol implementing mutual exclusion, working as follows: every process that wishes to access a critical resource draws a ticket, which is consecutively numbered. The process with the smallest number may access the resource first. If two processes draw a ticket concurrently, i.e., obtaining the same number, the process with the smaller process ID may access the resource first.
We monitored the following HyperLTL formula~\cite{conf/cav/FinkbeinerRS15}:
\begin{align*}
\forall \pi. \forall \pi'. &\LTLsquare (\mathit{sym}(\mathit{select}_\pi,\mathit{select}_{\pi'}) \wedge \mathit{pause}_\pi = \mathit{pause}_{\pi'})\\
&\rightarrow \LTLsquare (\mathit{pc}(0)_\pi = \mathit{pc}(1)_{\pi'} \wedge \mathit{pc}(1)_\pi = \mathit{pc}(0)_{\pi'}) \enspace ,
\end{align*}
where $\mathit{select}$ indicates the process ID that runs in the next step and $\mathit{pause}$ indicates whether the step is stuttering. Each process $i$ has a program counter $\mathit{pc}(i)$ and when process $i$ is selected, $\mathit{pc}(i)$ is executed. $\mathit{sym}(\mathit{select}_\pi,\mathit{select}_{\pi'})$ states that process $0$ is selected on trace $\pi$ and process $1$ is selected on trace $\pi'$.
Unsurprisingly, the implementation violates the specification, as it is provably impossible to implement a mutual exclusion protocol that is entirely symmetric~\cite{books/daglib/0080029}.
Figure~\ref{fig:bakery} shows the results of our experiment.
In this benchmark, we can observe that the language inclusion check, on which the trace optimization is based on, produces an overhead during the monitoring. Since the traces differ a lot, the trace analysis cannot prune enough traces to be valuable. As there are only a few instances (in this case 4), the trie optimization outperforms the previous version of $\tool$ massively on such a low instance count. The specification analysis, however, is always a valuable optimization.

\subsection{Case Study: Detecting Spurious Dependencies in Hardware Designs}






While HyperLTL has been applied to a range of domains, including security and
information flow properties, we focus in the following on a classical verification
problem, the independence of signals in hardware designs. We demonstrate how
$\tool$ can automatically detect such dependencies from traces generated from
hardware designs.
%


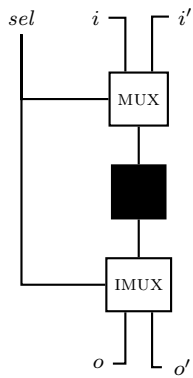
\begin{figure}[t]
  \centering
  \begin{tikzpicture}[auto,>=stealth',shorten >=1pt,thick,transform shape, scale=1]
  
  \node[draw,minimum size=20pt] (mux) {\textsc{mux}};
  \node[draw,fill,below=0.5 and -0.3 of mux,minimum size=20pt] (bb) {};
  \node[draw,below=0.5 and -0.3 of bb,minimum size=20pt] (imux) {\textsc{imux}};
  
  \node[above left=0.5 and 0 of mux] (i) {$i$};
  \node[above right=0.5 and 0 of mux] (iPrime) {$i'$};
  \node[left=0.5 of i] (sel) {$\mathit{sel}$};
  
  \node[below left=0.5 and -0.1 of imux] (o) {$o$};
  \node[below right=0.5 and -0.1 of imux] (oPrime) {$o'$};  
  
  \coordinate[xshift=-5pt] (muxi1) at (mux.north);
  \coordinate[xshift=5pt] (muxi2) at (mux.north);
  
  \coordinate[xshift=-5pt] (imuxo1) at (imux.south);
  \coordinate[xshift=5pt] (imuxo2) at (imux.south);
  
  \draw (i) -| (muxi1)
        (iPrime) -| (muxi2)
        (sel) |- (mux)
        (mux) -- (bb)
        (bb) -- (imux)
        (sel) |- (imux)
        (imuxo1) |- (o)
        (imuxo2) |- (oPrime)
        ;

\end{tikzpicture}
  \caption{\textsc{mux} circuit with black box.}
  \label{fig:example-circuit}
\end{figure}

\paragraph{Input \& Output.}

The input to $\tool$ is a set of
traces and a $\hyperltl$ formula. For the following
experiments, we generate a set of traces from the Verilog description of several example circuits by random
simulation. If a set of traces violates the specification,
$\tool$ returns a counter example.

\paragraph{Specification.}
We consider the problem of detecting whether input signals influence output signals in hardware designs.
We write $\vec i \ninfluences \vec o$ to denote that the inputs $\vec i$ do not influence the outputs $\vec o$.
Formally, we specify this property as the following $\hyperltl$ formula:
\begin{equation*}
\forall \pi_1 \forall \pi_2 \ldot
(\vec{o}_{\pi_1} = \vec{o}_{\pi_2}) \WUntil (\overline{\vec{i}}_{\pi_1} \neq \overline{\vec{i}}_{\pi_2}) \enspace,
\end{equation*}
where $\overline{\vec{i}}$ denotes all inputs except $\vec i$.
Intuitively, the formula asserts that for every two pairs of execution traces $(\pi_1,\pi_2)$ the value of $\vec{o}$ has to be the same until there is a difference between $\pi_1$ and $\pi_2$ in the input vector $\overline{\vec{i}}$, i.e., the inputs on which $\vec o$ may depend.

\paragraph{Sample Hardware Designs.}
\begin{figure}[t]
		\begin{lstlisting}[style={verilog-style},xleftmargin=.03\textwidth]
module counter(increase,decrease,overflow);
input increase;
input decrease;
output overflow;

reg[2:0] counter;

assign overflow = (counter==3'b111 
 && increase && !decrease);

initial
begin
  counter = 0;
end
always @($global_clock)
begin
if (increase && !decrease)
  counter = counter + 1;
else if (!increase && decrease
         && counter > 0)
  counter = counter - 1;
else 
  counter = counter;
end
endmodule
		
		\end{lstlisting}
		
	\caption{Verilog description of Example~\ref{ex:counter} (counter).}
	\label{fig:counter-verilog}
\end{figure}
We apply $\tool$ to traces generated from the following hardware designs.
Note that, since $\tool$ observes traces and treats the system that generates the traces as a black box, the performance of $\tool$ does not depend on
the size of the circuit.

\begin{table*}
	\caption{Results of $\tool$ on traces generated from circuit instances. Every instance was run 10 times with different seeds and the average is reported. Prototype refers to the first version of RVHyper~\cite{conf/tacas/FinkbeinerHST18} and RVHyper to the current implementation including the trie optimization.}
	\label{tbl:rvhyper-results}
	\centering
	\begin{tabular}{llllrrrr}
		\hline \noalign{\smallskip}
		instance & property & satisfied & \#\,traces & \multicolumn{2}{c}{prototype} & \multicolumn{2}{c}{RVHyper} \\
		&&&&  time & \#\,instances &  time & \#\,instances \\
		\noalign{\smallskip}\hline\noalign{\smallskip}
		\textsc{xor} & $i_0 \ninfluences o_0$ & no & 18 & 12ms & 222 & 6ms & 18  \\
		\textsc{xor} & $i_1 \ninfluences o_0$ & yes & 1000 & 16\,913ms & 499\,500 & 1613ms &  127 \\
		counter & $\textit{incr} \ninfluences \textit{overflow}$ & no & 1636  & 28\,677ms & 1\,659\,446 & 370ms & 2 \\
		counter & $\textit{decr} \ninfluences \textit{overflow}$ & no & 1142  & 15\,574ms & 887\,902 & 253ms & 22\,341 \\
		\textsc{mux} & $\vec i' \ninfluences \vec o$ & yes & 1000 & 14\,885ms & 49\,9500 & 496ms & 32 \\
		\textsc{mux2} & $\vec i' \ninfluences \vec o$ & no & 82 & 140ms & 3704 & 27ms & 1913 \\ \hline
	\end{tabular}
\end{table*}

\begin{example}[\textsc{xor}]
  As a first example, consider the \textsc{xor} function $\vec{o} = \vec{i} \oplus \vec{i}'$.
  In the corresponding circuit, every $j$-th output bit $o_j$ is only influenced by the $j$-the input bits $i_j$ and $i'_j$.
\end{example}

\begin{example}[\textsc{mux}]
This example circuit is depicted in Figure~\ref{fig:example-circuit}.
There is a black box combinatorial circuit, guarded by a multiplexer that selects between the two input vectors $\vec i$ and $\vec i'$ and an inverse multiplexer that forwards the output of the black box either towards $\vec o$ or $\vec o'$.
Despite there being a syntactic dependency between $\vec o$ and $\vec i'$, there is no semantic dependency, i.e., the output $\vec o$ does solely depend on $\vec i$ and the selector signal.

When using the same example, but with a sequential circuit as black box, there may be information flow from the input vector $\vec i'$ to the output vector $\vec o$ because the state of the latches may depend on it.
We construct such a circuit that leaks information about $\vec{i}'$ via its internal state.

The left part of Fig.~\ref{fig:runtime-comparison-prob} shows the total runtime of $\tool$ with the different optimizations and a combination thereof.
As observed in our previous experiments, the specification analysis, if applicable as in this case, is a valuable optimization consistently reducing the runtime and does so also when combined with the trace analysis.
As expected, the runtime is halved by exploiting symmetry and reflexivity in the formula.
From the plot we can also infer that the trace analysis is effective in a context with a majority of redundant traces.
For such a highly redundant setup the trace analysis reduces the overall runtime of the monitoring algorithm by several magnitudes.
With a decrease of similarity and redundancy in the traces the positive effect of the trace analysis steadily decreases up until the overhead of the trace analysis itself gets noticeable.
The decrease in runtime for configurations without trace analysis, which comes with reduced traces similarity, is explained by the fact that the more the input of the monitored traces is different the earlier trace tuples can get pruned as they satisfy the specification and thereby reduce the computational burden of the algorithm.
This is also the reason why the configurations with trace analysis show decreasing runtime behavior again as soon as the aforementioned effects dominate the runtime characteristics of the monitoring approach.
In contrast to that, the trie optimization provides a stable improvement on the running time.

\end{example}

\begin{example}[counter]
	\label{ex:counter}
	Our last example is a binary counter with two input control bits $\mathit{incr}$ and $\mathit{decr}$ that increments and decrements the counter.
	The corresponding Verilog design is shown in Figure~\ref{fig:counter-verilog}.
	The counter has a single output, namely a signal that is set to one when the counter value overflows.
	Both inputs influence the output, but timing of the overflow depends on the number of counter bits.
\end{example}

\paragraph{Results.}
The results of multiple random simulations are given in Table~\ref{tbl:rvhyper-results}.
Even the previous version of $\tool$ was able to scale up to thousands of input traces with millions of monitor instantiations.
The novel implemented optimization of $\tool$, i.e., storing the traces in a prefix tree data structure combined with our specification analysis, results in a remarkable speed-up.
Especially interesting is the reduction of the number of instances in the counter example. As there is only one input, the traces collapse in our trie data structure.
For the two instances where the property is satisfied (\textsc{xor} and \textsc{mux}), $\tool$ has not found a violation for any of the runs.
For instances where the property is violated, $\tool$ was able to find counter examples.

\section{Conclusion}

$\tool$ monitors a running system for violations of a HyperLTL specification.
We have introduced a novel trace storage optimization, based on a prefix-tree data structure, to existing optimizations implemented in $\tool$.

We demonstrated the impact of the optimizations on $\tool$s performance on several benchmarks of runtime verification problems. By providing a use case on how $\tool$ can be used to detect spurious dependencies in hardware design, we showed how $\tool$ can be used outside of classical security monitoring problems.
The functionality of $\tool$ thus complements model checking tools for HyperLTL, like MCHyper~\cite{conf/cav/FinkbeinerRS15}, tools for satisifability checking, like EAHyper~\cite{conf/cav/FinkbeinerHS17}, and tools for synthesis, like BoSyHyper~\cite{conf/cav/FinkbeinerHLST18}.

$\tool$ is in particular useful during the development of a $\hyperltl$ specification, where it can be used to check the HyperLTL formula on sample traces without the need for a complete model.
Based on the feedback of the tool, the user can refine the HyperLTL formula until it captures the intended policy.

In our current approach, the trace analysis and the trie representation are separate optimizations that cannot be applied at the same time. The integration of the two optimization is an interesting challenge for future work.

\bibliographystyle{splncs04}
\bibliography{main}

\end{document}